\documentstyle[psfig,prb,aps]{revtex}

\begin{document}

title{{\it Ab Initio} studies of the atomic structure and electronic
density
of states of pure and hydrogenated {\it a}-Si
}
\author{Ariel A. Valladares\cite{av} and Fernando Alvarez}
\address{Instituto de Investigaciones en Materiales, UNAM, Apartado Postal
70-360, M\'exico, D. F., 04510, MEXICO\\
}
\author{Zhihua Liu}
\address{Molecular Simulations, Inc., 9685 Scranton Road, San Diego, CA,
92121,
USA\\
}
\author{Juergen Stitcht}
\address{SciCo Inc., 5031 Palermo Dr., Oceanside, CA, 92057, USA\\
}
\author{John Harris}
\address{Institut fuer Festkoeperforschung, Forschungszentrum, Juelich GmbH,
D-52425, Juelich, Germany
}
\date{\today}
\maketitle

\begin{abstract}

We propose a method to simulate {\it a}-Si and {\it a}-Si:H using an {\it ab
initio} approach based on the Harris functional and thermally amorphisized
periodically continued cells with at least 64 atoms, and calculate their
radial
distribution functions.  Hydrogen incorporation was achieved via diffusive
random addition.  The electronic density of states (DOS) is obtained
using density functional theory with the aid of both the Harris-functional
and
Kohn-Sham-LDA approaches.  Two time steps are used, 2.44 and 10 fs for the
pure,
and 0.46 and 2 fs for the hydrogenated, to see their effect on the
topological
and DOS structure of the samples.  The calculated long time-step radial
features of {\it a}-Si are in very good agreement with experiment whereas
for
{\it a}-Si:H the short time-step partial and total radial features agree
well;
for the long time-step simulation molecular hydrogen appears during
annealing.
The long time-step {\it a}-Si has a well defined gap with two dangling
bonds,
that clears and increases upon hydrogen addition and relaxation, as
expected.
The short time-step structures have more defects, both dangling and floating
bonds, that are less characteristic of a good sample; however the radial
structures of {\it a-}Si:H are in better agreement with experiment
indicating
that the experimental work was done on defective samples.

\end{abstract}

\bigskip

(PACS: 71.23.Cq, 71.15.Pd, 71.55.Jv, 73.61.Jc)

\section{Introduction}

The scientific and technological relevance of {\it a}-Si, pure and
hydrogenated,
is well known and needs not be emphasized here; early work on the atomic
structure of the pure amorphous material started more than four decades ago,
both experimentally and theoretically.  Experimentally, work on the
amorphous phases of pure germanium and silicon evolved in parallel,
beginning
with the electrolytic approach of Szekely \cite{szeke} in 1951 and the
pioneering efforts of Richter and Breitling \cite{ribre} in 1958.
Theoretically, the first atomic models of both {\it a}-Si and {\it a}-Ge
appeared in the literature over thirty years ago. Grigorovici and Manaila
\cite{griman} and Coleman and Thomas \cite{coltho} in 1968 suggested
structures
based on arrangements of closely packed simplified Voronoi polyhedra that
have
the shape of truncated tetrahedra both eclipsed and staggered $60^\circ$
about
their common bond, leading to rings of five atoms and to boatlike rings of
six
atoms, as in the carbon and silicon amorphous clusters that we have recently
studied. \cite{valla}  As is well known, fivefold symmetry is
non-crystallographic and therefore yields an amorphouslike diffraction
pattern
with broad maxima.  The eclipsed configuration with a fivefold symmetry
structure is energetically unfavorable in the crystalline phases but occurs
in the amorphous form since atomic arrangements with a large internal energy
can appear in such frozen structures. \cite{mottdavis}

It was soon realized that such pure structures were incapable of being doped
with donors or acceptors, due to the presence of dangling bonds that masked
the
existence of an energy gap in the density of states (DOS) spectrum, and the
search began to find a solution to this problem.  Spear and Lecomber
\cite{spear} discovered that if amorphous silicon is grown in the presence
of
hydrogen the naturally existing dangling bonds passivate revealing, in this
manner, the presence of the gap. It was then possible to introduce dopants
whose influence was clearly seen.  This radically transformed the study of
amorphous semiconductors from an academic subject to a technologically
relevant
one that has led to the design of devices that are presently used in many
applications. In this manner the study of hydrogenated amorphous silicon
flourished \cite{street} and became one of the most important subjects in
the
area of amorphous semiconductors.

All these materials have been experimentally produced and characterized
without
theoretical guidelines that atomistic simulations may provide. In
particular,
several experimental radial distribution functions (RDFs) have been obtained
for
pure amorphous silicon, and some for hydrogenated amorphous silicon, using a
variety of techniques from X-ray to neutron diffraction, in order to better
comprehend their atomic structure. A consistent picture has emerged for the
structure of {\it a}-Si.  However, for {\it a}-Si:H the neutron diffraction
experiments needed to probe the hydrogen presence are relatively more recent
\cite{ivan} and the information is scarce; nevertheless some interesting
structural features have been identified for this system, which makes it
particularly appealing for theoretical simulations.

It is clear that, in principle, a full theoretical description of the
properties
of an amorphous solid depends on a complete knowledge of the atomic
structure.
However, for these solids there are an infinite number of possible
structures
and the best we can do in order to characterize their atomic arrangement is
to
use the RDF, also known as pair distribution function.  Defined as
$4\pi{r^2}\rho(r)dr$, where $\rho(r)$ is the density of particles at point
$r$,
the RDF gives the average number of pairs of atoms separated by a distance
lying
between $r$ and $r+dr$.  A given structure generates a well defined RDF but
a
given RDF corresponds to many atomic structures; that is, there is not a one
to
one correspondence between a given structure and its corresponding RDF.  For
this reason, clusters with a particular local structure have been used to
simulate the amorphous bulk and this has shed light on the effect of
specific
atomic arrangements on the electronic properties of these
materials.\cite{valla}

The experimental work of Refs. 10 to 16 for {\it a}-Si and Refs. 17 and 18
for
{\it a}-Si:H will be considered in this paper.  For hydrogenated silicon,
Ref.
18 reports the only complete study of the total RDF, together with the
partials
for Si-Si, Si-H and H-H.

On the theoretical side, a considerable amount of work has been done to
simulate
the atomic structure of {\it a}-Si, while the atomic structure of {\it
a}-Si:H
has been the subject of more recent work. A knowledge of their atomic
topology leads to better and more realistic calculations of the electronic
structure of these systems.  These efforts can be classified via two extreme
types: i) calculations that are carried out in samples that are constructed
essentially ``to order'' by switching bonds and adjusting dihedral angles,
and
that use {\it ad hoc} classical, parameter-dependent potentials constructed
for the specific purpose of describing them;  ii) quantum methods,
parameterized
and {\it ab initio}, that can deal from the outset with the thermalization
processes that generate the amorphous structure, and the study of their
corresponding electronic properties.  In between, one finds a variety of
hybrid
approaches.  In the present work we shall not consider the beautiful,
computationally cost-effective, work that has been done with classical
techniques in very large supercells of {\it a}-Si that contain tens of
thousands
atoms (see for instance Ref. 19 and work cited therein), since such methods
are
not transferable to other amorphous materials. Moreover, it is necessary to
generate new {\it ad hoc} parameters and potentials in order to deal with
new
materials, either from experimental results or from {\it ab initio}
calculations.  Furthermore, it has been established \cite{rog} that even
though
some of these potentials describe the RDF, they fail in differing degrees to
describe other properties of the amorphous lattice, and this is
understandable
since no matter how good a classical approach may simulate the atomic
interactions, there will be quantum properties beyond the capability of
these
techniques.

Quantum methods answered some of the unsolved questions left by the
classical
approaches, and were themselves of several kinds. For example, there has
been
some interesting work done using tight binding methods for pure silicon,
\cite{kwon} where a transferable model is found by fitting the energies of
silicon in various bulk crystal structures and examining functional
parameterizations of the tight binding forms.  For hydrogenated silicon a
transferable model has also been found by fitting it to silane. \cite{li}
On
the other hand, there are {\it ab initio} methods that attempt to answer all
the
questions from first principles and are generally applicable without
adjustment
of parameters, but are very demanding in computer resources and so are
limited
to handling relatively small amorphous cells.  The question is, how
generally
can an {\it ab initio} method that uses a reasonably sized supercell be
applied,
and how accurate can one expect the results to be?  The present work
addresses
these issues for amorphous silicon and for hydrogenated amorphous silicon.

\section{Antecedents}

For our purposes, the theoretical calculations described in Car and
Parrinello
\cite{car1} through to Lee and Chang\cite{lee} for {\it a}-Si and in Buda
{\it et al.}\cite{buda} through to Tuttle and Adams\cite{tuttle} for {\it
a}-
Si:H are appropiate since they approach the structural problem by generating
amorphous cells using first-principles quantum methods.

More than a decade ago Car and Parrinello\cite{car1} performed the first
{\it ab
initio} molecular dynamics (MD) study in an fcc cell with 54 atoms of
silicon
using their plane wave MD method.  In their approach a non-local
pseudopotential
was used together with the parameterized local density (LDA) form of Perdew
and
Zunger for the exchange-correlation effects.  They obtained good agreement
with
the experimental RDF of {\it a}-Ge, rescaled to simulate {\it a}-Si, up to
the
second radial peak and argued that because of the size of the cell used
distances larger that 6 \AA\ could not be studied.  They pointed out that
the
comparison of the simulated atomic structure to the experimental ones should
be
done with care since a large number of defects were found in their results.
The
simulation was started above the melting point, at about 2,200 K, and then
the
liquid was allowed to evolve for $\approx 0.7\times10^{-12}$s before it was
quenched down to $\approx$ 300 K at a cooling rate of $\approx 2\times
10^{15}$
K/s. During the initial quenching the volume of the cell was gradually
increased
to 1080 \AA$^3$, the crystalline value.  This technique of quenching from
the
melt has been used in subsequent work although handling the transition from
the
liquid to the amorphous phase is not an easy task since a volume change has
to
be dealt with and because liquid silicon is metallic with an average
coordination number of between 6 and 7, the quenching preserves some of this
overcoordination.

For example, Drabold {\it et al.}\cite{drabold} use a density functional
theory,
local density approximation (DFT-LDA), molecular dynamics approach developed
by
Sankey {\it et al.}\cite{san} based on a simplification of the Kohn-Sham
equations as developed by Harris,\cite{harris} and starting with a 64-atom
simple cubic cell in the diamond structure with one vacancy, they generate
an
``incompletely melted'' sample by heating it up to 8,000 K.  The free
evolution
of the cell then results in the system acquiring a highly disordered
liquidlike structure before final quenching to a solid.  They state that
their results for the RDF agree well with experiment, without making a
detailed,
direct comparison.  They find only four coordination defects, two dangling
and
two floating bonds, for appropiate values of the rate of free evolution of
the
cell.  After annealing at 300 K only two defects survived.

A more complete report of the Car and Parrinello results is given in Stich
{\it
et al.},\cite{stich} where a cooling rate from the melt of $10^{14}$ K/s is
used. This slower cooling rate seems to be sufficient to recover a
tetrahedral
network that nevertheless contains several coordination defects as well as a
large fraction of distorted bonds.  Annealing at 900 K reduces the defects
and
the RDF they obtain has two peaks that seem to agree with the first two
peaks of
experiments.  The study performed by Drabold {\it et al.} was extended
\cite{fedders1} to the 216 atom periodic supercell of Wooten, Winer and
Weaire
\cite{woo} and a more complete analysis of the relationship of structural
defects, spectral defects and interatomic distances was carried out.
Lee and Chang\cite{lee} perform {\it ab initio} simulations on a 64-atom
silicon cell and quench incompletely melted samples as in Ref. 24, but they
find that the third peak of the RDF practically disappears, and that more
dangling and floating bonds occured than in previously generated samples
from
liquid-quenched simulations, and than in the samples generated by Drabold
{\it et al.} \cite{drabold}

Theoretical work on {\it a}-Si:H has been less abundant because the
experimental
RDF results are limited and it is more difficult to model interactions of H
and
Si and cope with the  time steps needed for computer simulations of hydrogen
diffusion.  In addition, the strong dependence of the {\it a}-Si:H structure
on
deposition conditions has to be reflected in the simulations, together
with the chemical reactivity of hydrogen and its zero point energy; all this
requires the use of quantum mechanical methods.  The {\it ab initio} work
performed for this system is found in Refs. 29 to 31.  In some of the
models H is incorporated into a previously created amorphous network of pure
{\it a}-Si by hand, \cite{fedders2} with the consequent inhibition of
hydrogen
diffusion, whereas Buda {\it et al.} \cite{buda} and Tuttle and Adams
\cite{tuttle} do allow bulk diffusion of H by first creating a liquid sample
of
the material and then quenching it, a procedure that has known shortcomings,
such as overcoordination, no definite gap, etc.

Specifically, the plane wave MD Car-Parrinello method was applied to
amorphous
hydrogenated silicon by Buda {\it et al.}\cite{buda} using a cubic cell of
64
silicons and 8 hydrogens (11\% concentration).  They start out with a liquid
material containing both silicon and hydrogen atoms that is rapidly
quenched, maintaining a density equal to the  value of the crystalline
material.
They report only partial distribution functions and, as is usually the case,
the
H-H RDF is poorly reproduced and is not compared to the existing
experimental
neutron diffraction data.  The DFT-LDA approach of Sankey and coworkers was
applied to this material by Fedders and Drabold\cite{fedders2} using several
cells based on a two-defect 63 silicon atom supercell that was constructed
in
previous work.\cite{drabold,fedders1} The hydrogen was introduced into the
amorphous silicon sample by hand so that the H-atoms are located near (1.5
\AA)
the corresponding dangling bond.  To eliminate the strained bond defects
they removed the silicon considered to be the center of the
strain defect and put H atoms near the 4 dangling bonds.  Posterior
relaxation
allowed the hydrogens to be trapped by the dangling bonds.  Fedders and
Drabold\cite{fedders2} do not report any RDF, total or partial.  Tuttle and
Adams \cite{tuttle} also use the Harris functional in the DFT-LDA code
developed
by Sankey {\it et al.} and apply it to a cell of 242 atoms with 11\%
hydrogen.
They generate the structure from a liquid at $\approx$ 1800 K and quench it
to produce an amorphous structure at 300 K.  Since they were not concerned
with
real time dynamics, they set the mass of hydrogen equal to the mass of Si,
thereby allowing the use of a large time step in the annealing process (4
fs).
However, this means that the RDFs could not reflect real diffusion processes
of the hydrogens in the cell, and correspondingly only the Si-Si and the
Si-H
RDFs are reported.  A significant percentage of defects with only 90\% of
the
silicon being fourfold coordinated is found.

Even though tight binding calculations have been labeled as ``highly
arbitrary
and inadequate'', \cite{fedders1} some of them deserve special mention.
Biswas
{\it et al.} \cite{biswas} using this approach studied the electronic
structure
of dangling and floating bonds in amorphous silicon.  Fedders \cite{fedders}
looked into the energetics of defects and found that tight binding gives
surprisingly good results for the energy eigenvalues and the degree of
localization of the defect states if some radial dependence is included in
the
hopping matrix elements.  Colombo and Maric \cite{colo} reported the first
tight-binding molecular-dynamics simulation of the defect-induced
crystal-to-amorphous transition in crystalline silicon.  In particular, two
recent ones are specially relevant for the present work: Yang and
Singh\cite{yang} find that the total average energy per silicon atom is a
minimum when the hydrogen concentration is in the range 8-14\%, the optimum
experimental range found; Klein {\it et al.} \cite{frau} report a H-H RDF
closer to experimental results than previous work.  Both studies start out
from
liquid samples that are fast quenched to generate the hydrogenated amorphous
silicon and therefore, as in all other cases, a large percentage of defects,
floating or dangling bonds, are naturally created.  For a recent account of
the
state of the art of tight binding methods see Ref. 39 where both
parameterized and {\it ab initio} approaches are considered.

\section{Method}

In this work we report the generation of samples of both {\it a}-Si
and {\it a}-Si:H using a new approach (with four different time steps) that
leads to structures with a minimum of coordination defects.  We use {\it
FastStructure}, \cite{fast} a DFT code based on the Harris functional, and
optimization techniques based on a fast force generator to allow simulated
annealing/molecular dynamics studies with quantum force calculations.
\cite{harris2}  We use the LDA parameterization due to Vosko, Wilk and
Nusair
(VWN) \cite{vosko} in the simulations.  The core is taken as full which
means
that an all electron calculation is carried out, and for our simulations a
minimal basis set of atomic orbitals was chosen with a cutoff radius of 5
\AA,
(compare to values of $\approx2.6$ \AA\ used by Sankey {\it et al.}), a
compromise between cost and accuracy.  The physical masses of hydrogen and
silicon are used throughout and this allows us to see realistic diffusive
processes of the hydrogen atoms in the supercell.  The default time step is
given by $\sqrt{m_{min}/5}$, where $m_{min}$ is the value of the smallest
mass
in the system; however, in order to better simulate the dynamical processes
that
occur in the amorphisation a time step of 10 fs (compare to 2.44 fs, the
default
value) was also used for the pure silicon and 2 fs was used when hydrogen
was
introduced, in addition to 0.46 fs, the default time step.  The forces are
calculated using rigorous formal derivatives of the expression for the
energy
in the Harris functional, as discussed by Lin and Harris.\cite{lin}  The
evaluation of the 3-center integrals that contribute to the  matrix elements
in the one-particle Schrodinger equation is the time-limiting feature of
{\it
FastStructure} and each is performed using the weight-function method of
Delley. \cite{dell}

Since it is clear that quenching from a melt or from partially melted
samples
generates undesirable structures, we took a different path.  Our process,
like
the ones mentioned above, is not designed to reproduce the way an amorphous
material is grown, but has the objective of generating an amorphous sample
that
would represent adequately the ones obtained experimentally. We amorphisized
the
crystalline diamond structure with 64 silicon atoms in the cell (a
crystalline
density of 2.33 g/cm$^3$) by slowly heating it from room temperature to well
above the glass transition temperature and just below the melting point, and
then slowly cooling it to 0 K.  This then was followed by cycles of
annealing
and quenching at temperatures suggested by experiment, to let the structure
obtained adjust to the local energy minimum.  Heating, cooling, annealing
and
quenching cycles were carried out using either the 2.44 fs or the 10 fs time
step throughout.

To treat {\it a}-Si:H we first constructed two amorphous silicon structures
as
follows: One was created as described above, with a time step of 2.44 fs,
maintaining the crystalline density, and then expanded to reproduce the
experimental density, 2.2 g/cm$^3$ of the hydrogenated structure, once
hydrogens are incorporated. The other was created using a time step of 10 fs
from an expanded crystalline cell so that the density of the hydrogenated
cell
(2.2 g/cm$^3$) corresponds to the reported experimental values.  Once these
expanded pure amorphous silicon cells were constructed, we introduced 12
hydrogen atoms at the following relative cell positions: (1/4, 1/4, 1/4),
(3/4, 1/4, 1/4), (3/4, 3/4, 1/4), (1/4, 3/4, 1/4), (1/2, 1/4, 1/2),
(3/4, 1/2, 1/2), (1/2, 3/4, 1/2), (1/4, 1/2, 1/2), (1/4, 1/4, 3/4),
(3/4, 1/4, 3/4), (3/4, 3/4, 3/4), (1/4, 3/4, 3/4), and annealed them at
temperatures suggested by experiment.

Specifically, for {\it a}-Si the  cell was heated from 300 K to 1,680 K,
just
below the melting temperature of crystalline silicon, in 100 steps of 10 fs,
and
immediately cooled down to 0 K in 122 steps of 10 fs; this was done also for
a
time step of 2.44 fs.  The heating/cooling rate was $5.66\times10^{15}$K/s
for
2.4 and $1.38\times10^{15}$K/s for 10 fs, approximately.  The atoms were
allowed
to move freely within each cell of volume (10.8614 \AA)$^3$ with periodic
boundary conditions. Once this first stage was complete, we subjected each
cell
to annealing cycles at 300 K (below microcrystallization \cite{street}) with
intermediate quenching processes.

For {\it a}-Si:H we implemented two different procedures.  First, we used
the
amorphous pure silicon cell generated above with a time step of 2.44 fs and
then expanded it to a volume of (11.0620 \AA)$^3$.  Second, we amorphisized
a previously expanded crystalline cell of 64 silicon atoms with the same
volume
of (11.0620 \AA)$^3$ using a 10 fs time step and repeating the cycles
described
above.  We then placed the 12 hydrogens distributed throughout the
amorphous cells and subjected the system to annealing cycles using a time
step
of 0.46 fs for the first cell (the 2.44/0.46 cell), and 2 fs for the second
one
(the 10/2 cell).  The annealing cycles consist of two cycles of 50 steps at
300 K for the large time step sample and one cycle of 200 steps for the
small
time step sample, with in between quenches down to 0 K, to allow the
hydrogens
to diffuse and move in the cells.  This gives a concentration of hydrogen of
practically 16\%, adequate to compare with existing experimental results.

Once the atomic structures were constructed and their respective RDFs
obtained,
we analysed their electronic density of states at the $\Gamma$ point of the
Brillouin Zone using both {\it FastStructure} and the DFT approach of
Hohenberg
and Kohn\cite{hohenberg64} and Kohn and Sham\cite{kohn65} implemented in the
{\it ab initio DMol3} comercial code;\cite{dmol3} both codes were used to
calculate the energy levels and DOS curves of the amorphous atomic
structures
generated.  {\it DMol3} treats the electronic structure of periodic solids
by
solving the Kohn-Sham self-consistency equations within local or nonlocal
density approximations; in our work we carried out just energy calculations
of
the structures found with {\it FastStructure} and used the LDA
approximation.
We also used a double numerical basis set that includes {\it d}-polarization
of the atoms (DNP) and the frozen inner core orbitals approximation; a
medium
grid was used for the numerical calculations.  The SCF density parameter
that
specifies the degree of convergence for the LDA density was set at
$10^{-6}$.
The HOMOs, LUMOs and DOS curves were calculated since they are necessary to
study the behaviour of the forbidden energy gaps, the position of the Fermi
levels, and the gap levels introduced by either dangling bonds, floating
bonds
or hydrogen states.

\section{Results and discussion}

Defining a dangling or a floating bond is to some extent arbitrary since one
has
to choose some interatomic distance; Drabold {\it et al.}\cite{drabold}
chose an
interatomic distance of 2.7 \AA, although no clear justification for this
choice
is given. Lee and Chang\cite{lee} use the distance at which the minimum of
the
RDF occurs, 2.73 \AA, which leads to an average number of nearest neighbors
in
the amorphous cell of 3.9.  Were we to use the minimum value of the RDF of
Fedders {\it et al.},\cite{fedders1} $\approx  1.2\times2.35=2.82$ \AA, it
is
certain that the number of defects reported by Drabold and coworkers would
have
changed significantly.  An analysis of  structural defects {\it versus}
spectral
defects was performed by these authors and it was concluded that dangling
bonds
give rise to defects within the gap, that some strained tetrahedral
structures
may also generate gap states, but that it seems floating bonds do not create
gap
states.  Recent work, however, indicates that floating bonds may generate
gap states,\cite{ital} and this conclusion is also borne out in the results
reported here.

Generating a clean gap in such small cells is a difficult task since defects
always appear in the simulations; so it seems that the most one can hope for
is
to minimize the number of defects and so obtain a reasonably clean
electronic
gap.  On the other hand, generating RDFs, total and partial, that compare
favorably with experiment is very important as is reproducing the
hydrogen dynamics and its experimental behaviour of passivating the dangling
bonds in amorphous silicon. This 'cleans' the gap, in spite of the limited
size
of the cells.  Herein we address these issues and report total RDFs for
pure and hydrogenated amorphous silicon, and partial RDFs for
silicon-silicon,
silicon-hydrogen and hydrogen-hydrogen in the hydrogenated samples and also
report HOMOs, LUMOs and their DOS curves.

\subsection{The atomic structure}

For {\it a}-Si the RDF obtained from our simulation with 10 fs agrees very
well
with experiment from 0 to 10 \AA, Fig. 1.  In this work we fourier-smoothed
all
the RDF results to have adequate curves to allow comparison with experiment,
since the small number of atoms in the cell leads to statistical
fluctuations
that are not representative of the bulk.  In Fig. 1 we do a direct
comparison of
our results and the upper and lower experimental bounds from Refs. 10 to 16
and,
considering that not all experimental results cover the range 0 to 10 \AA,
the
agreement is excellent, including the existence and position of the third
and
the fourth amorphous peaks.  The evolution of the crystalline peaks can be
clearly observed in our simulations, Fig. 2, and we can say that the first
crystalline peak remains as the most prominent feature in the amorphous
material. The second crystalline peak also remains but is widened and highly
diminished since it contributes to filling in the first and second
crystalline
valleys. The third crystalline peak disappears to contribute to the valleys
between the second and third and the third and fourth crystalline peaks, and
the fourth crystalline peak disappears to contribute to the filling in the
valleys between the third and fifth crystalline peaks.   The fifth
crystalline
peak generates a third amorphous peak slightly displaced to smaller
distances.
This figure presents a comparison of the position of the crystalline peaks
and
the amorphous structure where all this can be better appreciated.

The total and partial RDF simulations for {\it a}-Si:H also compare very
favorably with the experimental results given in Ref. 18, where a more
complete
and more recent study is reported.  In Figs. 3 and 4 we show the comparison
of
the total RDFs and the partial H-H RDFs with experiment for the two cells,
2.44/0.46 (Fig. 3) and 10/2 (Fig. 4), where the peak due to the presence of
molecular hydrogen that appears in the simulation of the 10/2 cell has been
removed in order to better illustrate the agreement of our results with
experiment.  In particular, the H-H correlation function that we obtained
for
the 2.44/0.46 fs cell is close to the experimental results of Bellissent
{\it
et al.}\cite{belli} unlike previous {\it ab initio} simulations. As we shall
see later, it turns out that this sample contains more dangling and floating
bonds than the 10/2 fs cell.  Since the number of hydrogen atoms in the
cells
is smaller than the number of silicon atoms, a larger fourier smoothing for
the
H-H partial was used in order to compare our results with experiment.  In
Fig. 5 we are able to see the dynamics of the originally evenly distributed
hydrogen (see Fig. 6 for the distribution of hydrogens) since for
the 10/2 fs cell a peak appears in the total and H-H RDFs due to the
formation
of molecular hydrogen at an interatomic distance of  0.875 \AA, (see Ref. 49
where a molecular radius of 0.86 \AA\ for hydrogen in a crystalline silicon
cluster is reported).  No formation of molecular hydrogen shows up in the
2.44/0.46 fs cell since the large number of dangling bonds inhibits it.
Figs.
7 to 10 depict four structures that clearly indicate the passivation process
of
the existing dangling bonds in the pure amorphous silicon samples due to
hydrogenation: Fig. 7(9) shows 5(2) dangling bonds that existed in the pure
amorphous sample for the 2.44(10) fs cell, for a cutoff radius of
2.815(2.743)
\AA\ below which the number of neighbors is 4(4); Fig. 8(10) shows the
2.44/0.46(10/2) fs cells indicating that 3(1) bonds were passivated by
hydrogens, 2(1) by silicons, and 3(2) new ones appeared; here the same
cutoff
radius mentioned above for Si-Si was used, whereas the cutoff radius for
Si-H was 1.9(1.9) \AA\ which is the position of the valley minimum between
the
first and second peaks in the Si-H partial RDF (See Figs. 11 and 12).
The Si-Si and Si-H partial RDF also agree well with experiment, as indicted
in Figs. 11 and 12 where a direct comparison is shown.

The 2.44/0.46 cell shows a large number of floating bonds, 11, after the
hydrogen diffusion process, as opposed to 3 floating bonds (fb) for the 10/2
fs
cell.  Of the 11 fb, 6 are due to hydrogen atoms that bond to silicons in
addition to the existing tetrahedral coordination and it is for this cell
that
the addition of H decreases the size of the electronic gap, as will be shown
next, indicating the important role of this type of bonds in reducing the
gap.

\subsection{The electronic density of states}

Using {\it ab initio} methods in order to study electronic characteristics
of
amorphous materials invariably forces one to consider cells of some hundred
atoms at most, due to the limitations in computing power of present day
resources.  Therefore it is desirable to have practically no defects in
these
cells since then they would better represent what occurs in the best
extended
samples.  However, as stated before, simulations in cells of 64 to 216 atoms
usually have a large concentration of dangling and floating bonds.  That is
why
we have developed a different approach to generate amorphous cells that seem
to
give a smaller concentration of these defects and that are therefore more in
agreement with good bulk samples.

As mentioned above for {\it a}-Si our two amorphous cells 2.44/10 have 5 and
3
dangling bonds and 5 and 1 floating bonds for a cutoff radius of 2.763/2.772
\AA.  At these distances the number of nearest neighbors (area under the
first
amorphous peak) is equal to 4.  Calculations of the DOS curves, and HOMOs
and
LUMOs with both {\it FastStructure} and {\it DMol3} give clean gaps of 0.414
eV
and 0.173 eV respectively for the 2.44 fs cell and 0.744 eV and 0.385 eV
respectively for the 10 fs cell .  The gaps were obtained just by looking at
the HOMO and LUMO values; no attempt was made at sorting out the states near
the gap due to dangling and/or floating bonds.  Figs. 13 and 14 show the DOS
curves where the gaps are clearly indicated.  It should be kept in mind that
density functional calculations using the local density approximation tend
to
underestimate band gaps.

Two procedures were carried out in order to generate the amorphous cells of
hydrogenated silicon of the correct density; both imply an expansion of the
cell
either before the amorphisation or after, as described above.  When the cell
is
expanded after the amorphisation a larger number of defects is created, 6
dangling bonds and 2 floating bonds, for the cutoff radius of 2.763 \AA.
Then
12 hydrogens were placed as described above and indicated in Fig. 6 and the
cell was
annealed as described above.  It should be emphasized that in this method
the
hydrogen atoms were left free to diffuse, unlike other results reported in
the
literature where the hydrogens are placed in {\it ad hoc} positions to
saturate
the dangling bonds.  Figs. 15 and 16 show the DOS curves obtained with {\it
FastStructure} and {\it DMol3} for hydrogenated amorphous silicon, the gap
values are 0.324 eV and 0.142 eV for the 2.44/0.46 fs cell and 0.787 eV and
0.483 eV for the 10/2 fs cell, respectively. The result obtained with the
10/2
fs cell follows the trend found in experiments that the addition of hydrogen
increases the gap value with respect to pure silicon.  The result for the
2.44/0.46 fs cell indicate that the size of the gap can be diminished by the
presence of the 6 hydrogen-related and 5 silicon-related floating bonds;
this
supports the idea expressed recently by Fornary and coworkers. \cite{ital}

\section{Conclusions}

Dealing theoretically with a covalent amorphous material in bulk is
difficult
because there are many possible structures for a given RDF.  Continuous
random-network models have achieved considerable success in generating a-Si
structures but these models are not easily generalizable to other glassy
materials or systems with different species of atoms.  That is why the
simulation of covalent amorphous semiconductors in the bulk, both pure and
alloyed, has encountered serious obstacles.  In the present work, we
have been able to perform {\it ab initio} simulations of the atomic
structure
and
electronic features of amorphous pure and hydrogenated silicon and
simulation of their total and partial atomic RDFs that
reproduce many features shown by the experimental data. In particular,
our RDFs agree very well with the corresponding experimental distributions
up to the third and fourth peaks, as can be seen in the direct comparisons
made
in the figures.  In addition, the dynamics of atomic hydrogen,
including the formation of molecular hydrogen under certain conditions, and
the
passivation of dangling bonds was observed.  The electronic density of
states
of the
atomic distributions generated indicate the presence of a gap for the pure
amorphous silicon samples that widens when the sample is annealed after
hydrogen
is distributed within the cell and the number of defects (dangling
and floating bonds) is kept low.

The procedure that we used to generate these structures is different from
the
ones found in the literature and leads to amorphous samples with fewer
dangling bonds and fewer overcoordinated defects.  It consists of
heating a crystalline sample of 64 atoms of silicon just below its melting
temperature and then cooling it down to 0 K with subsequent annealing and
quenching cycles at temperatures dictated by experiment.

We believe that our procedure can be generalized to other monatomic or
diatomic
amorphous semiconducting materials and therefore allows  {\it ab initio}
techniques to be used in other areas to provide more representative
structures
and a more complete description of their atomic and electronic
characteristics
than is possible with parameterized methods.  Furthermore, {\it ab initio}
simulations can be used to provide the parameters needed in non {\it ab
initio}
techniques to be able to handle larger amorphous cells.

\bigskip

\begin{center}

{\bf ACKNOWLEDGMENT}

\end{center}

AAV thanks DGAPA-UNAM for financial support to spend a sabbatical year at
{\it Molecular Simulations, Inc.} where this work was begun and for
financial
support to carry out this research through project IN101798 ({\it Estructura
Electr\'onica y Topolog\'{\i}a At\'omica de Silicio Amorfo Puro y
Contaminado}).  FA thanks CONACyT for supporting his PhD studies.   This
work
was done on an Origin 2000 computer provided by DGSCA, UNAM.

\pagebreak

\begin{center}

FIGURE LIST

\end{center}

\bigskip
\bigskip

FIG. 1.  RDFs for {\it a}-Si.  The lighter lines are the experimental upper
and lower bounds (see text).  The dark line is our simulation where the
third and fourth peaks are well reproduced.

\bigskip

FIG. 2.  Evolution of the crystalline peaks during amorphisation.  The RDF
for
{\it a}-Si (dark line) has four well defined peaks that are generated by
several crystalline peaks (lighter vertical lines).

\bigskip

FIG. 3.  Direct comparison of the simulated (dark lines) total RDF for
{\it a}-Si:H and partial RDF for H-H (Inset) (with 15.79\% hydrogen
concentration) to the experimental (light lines) results for the short time
steps cell (2.44/0.46 fs).

\bigskip

FIG. 4.  Direct comparison of the simulated (dark lines) total RDF for
{\it a}-Si:H and partial RDF for H-H (Inset) (with 15.79\% hydrogen
concentration) to the experimental (light lines) results for the long time
steps cell (10/2 fs).  The peak due to molecular hydrogen has been removed
for clarity.

\bigskip

FIG. 5.  Total RDF for the 10/2 fs cell of hydrogenated amorphous silicon
(dark line) that shows the presence of molecular hydrogen, compared to the
experimental results (light line).

\bigskip

FIG. 6.  Starting positions of the 12 hydrogens placed within the amorphous
cells of pure silicon.  The relative coordinates are given in the text.

\bigskip

FIG. 7.  Short time step cell of pure amorphous silicon (2.44 fs) that
shows the existence of 5 dangling bonds.  A cutoff radius of 2.815 \AA\ was
used, below which the total number of nearest neighbors is 4.

\bigskip

FIG. 8.  When hydrogen is added to the short time steps cell, and
the sample annealed, 3 dangling bonds are passivated by
hydrogens and 2 by silicons but 3 new ones appear; however, 11 new floating
bonds are formed (not shown).

\bigskip

FIG. 9.  Long time step cell of pure amorphous silicon (10 fs) that
shows the existence of 2 dangling bonds.  A cutoff radius of 2.743 \AA\ was
used, below which the total number of nearest neighbors is 4.

\bigskip

FIG. 10.  When hydrogen is added to the long time steps cell, and
the sample annealed, all dangling bonds (2) are passivated, one
by hydrogen and one by silicon, and 2 new ones appear.  The number of
floating bonds now is only 3 (not shown).  The appearance of molecular
and atomic hydrogen is indicated.

\bigskip

FIG. 11.  Partial Si-Si and Si-H RDFs for the hydrogenated short time steps
cell.  The simulated results are represented by dark lines whereas the
experimental ones are represented with lighter lines.

\bigskip

FIG. 12.  Partial Si-Si and Si-H RDFs for the hydrogenated long time steps
cell.  The simulated results are represented by dark lines whereas the
experimental ones are represented with lighter lines.

\bigskip

FIG. 13.  DOS curves for the 2.44 fs cell of {\it a}-Si
calculated using {\it FastStructure}, curve (a), and {\it DMol3}, curve
(b).  HOMOs and LUMOs are shown that lead to a gap of 0.414 eV for (a) and
0.173 eV for (b).

\bigskip

FIG. 14.  DOS curves for the 10 fs cell of {\it a}-Si calculated using
{\it FastStructure}, curve (a), and {\it DMol3}, curve (b).  HOMOs and LUMOs
are shown that lead to a gap of 0.744 eV for (a) and 0.385 eV for (b).

\bigskip

FIG. 15.  DOS curves for the 2.44/0.46 fs cell of {\it a}-Si:H calculated
using
{\it FastStructure}, curve (a), and  {\it DMol3}, curve (b).  HOMOs and
LUMOs
are indicated and the gap values are 0.324 eV and 0.142 eV, respectively.

\bigskip

FIG. 16.  DOS curves for the 10/2 fs cell of {\it a}-Si:H calculated using
{\it FastStructure}, curve (a), and  {\it DMol3}, curve (b).  HOMOs and
LUMOs
are indicated and the gap values are 0.787 eV and 0.483 eV, respectively.

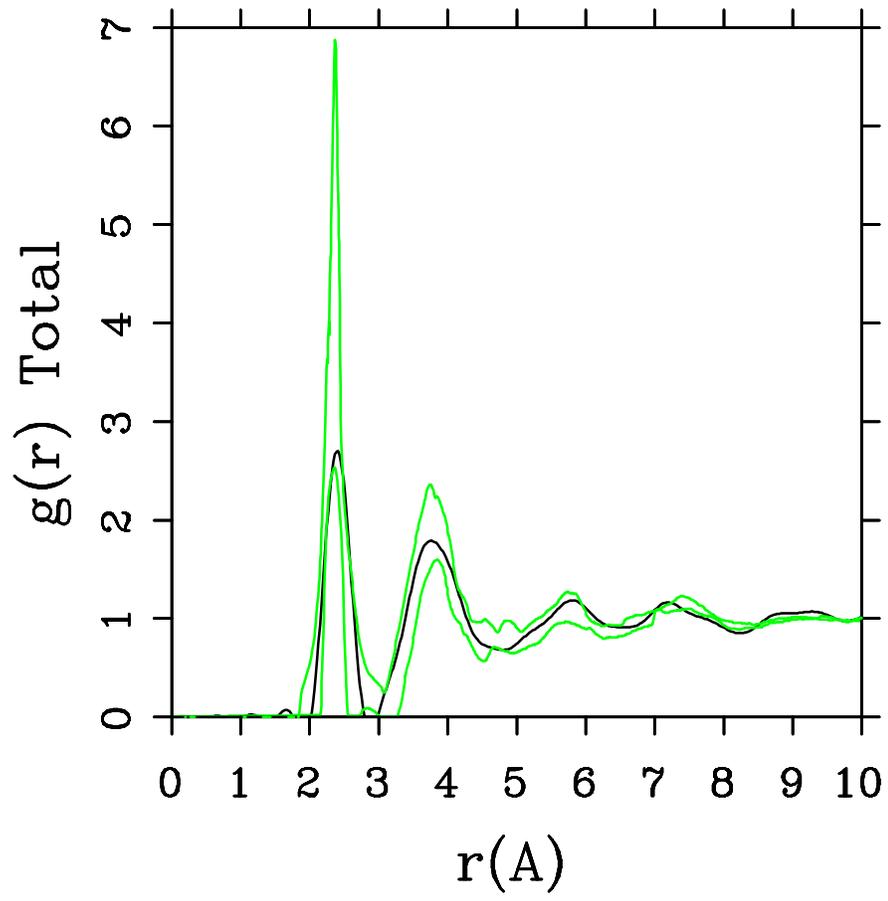
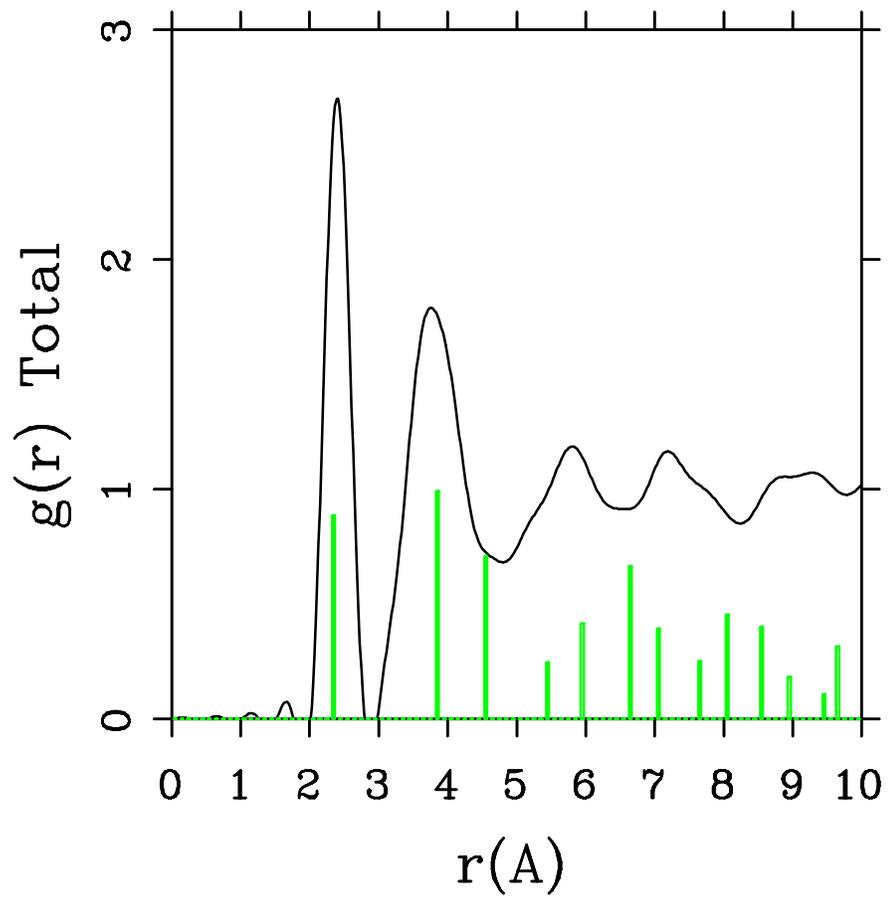
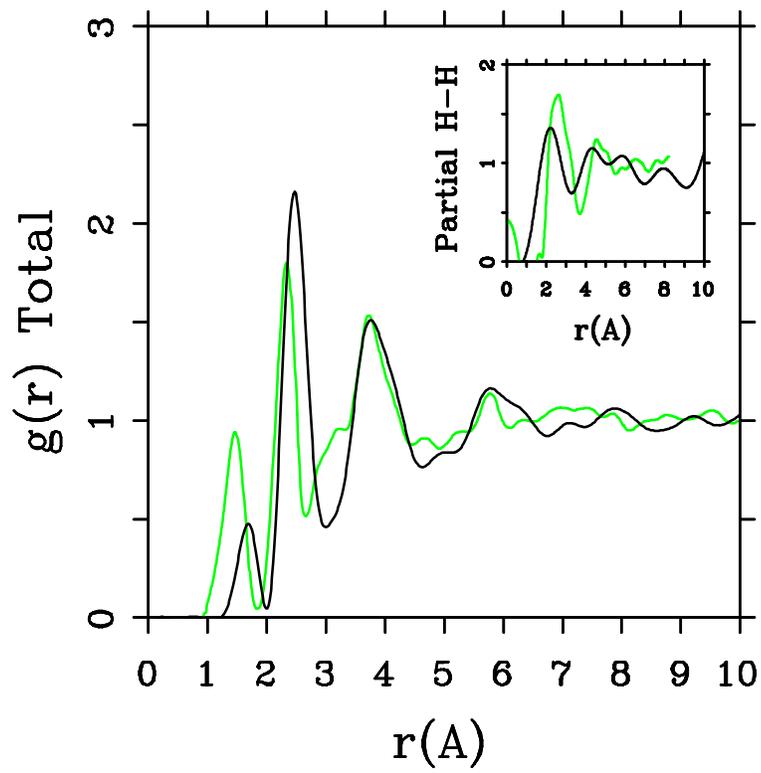
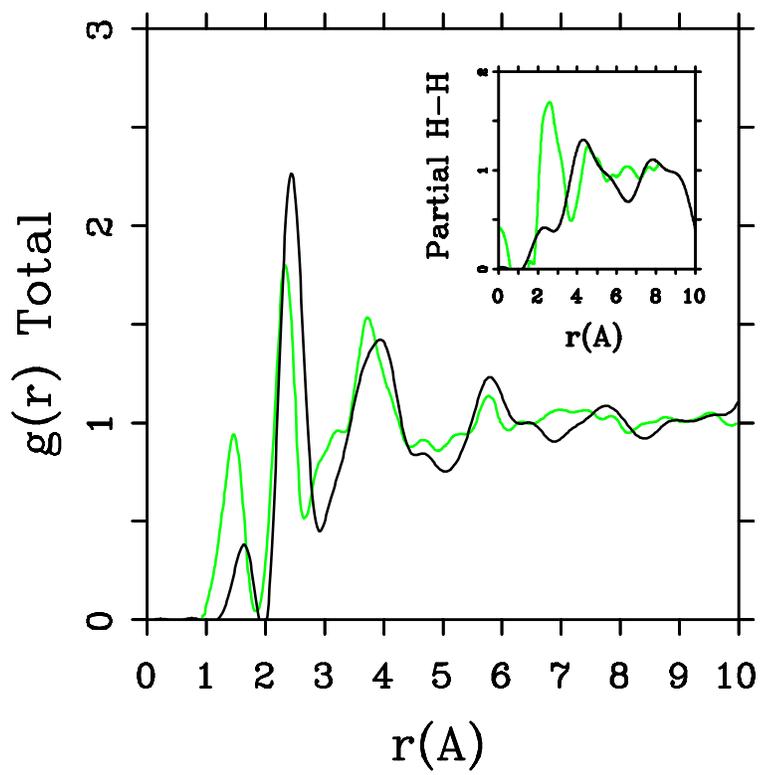
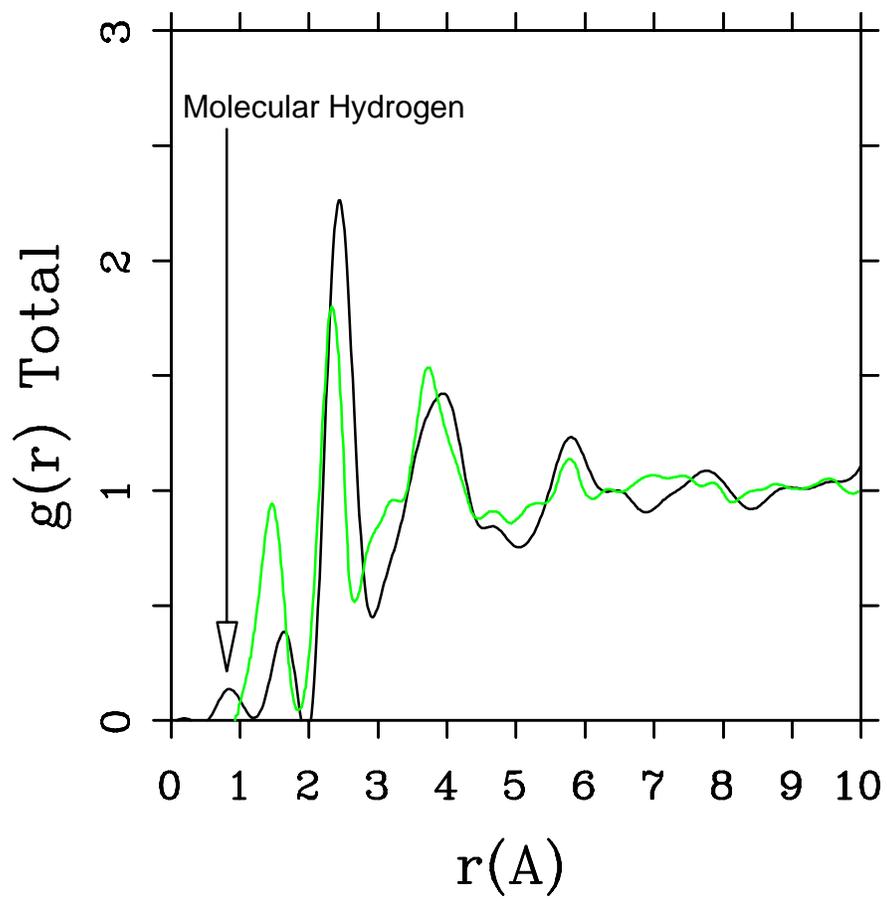
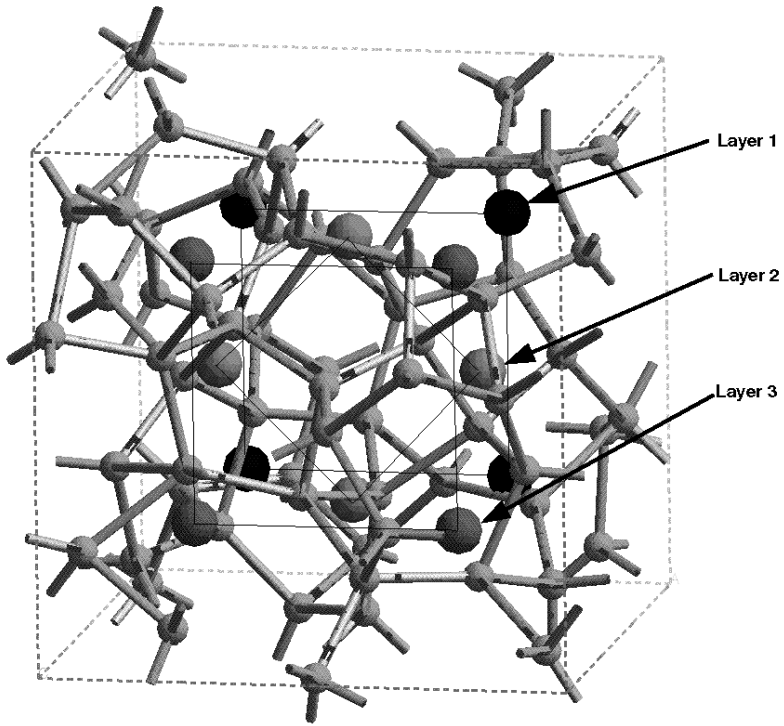
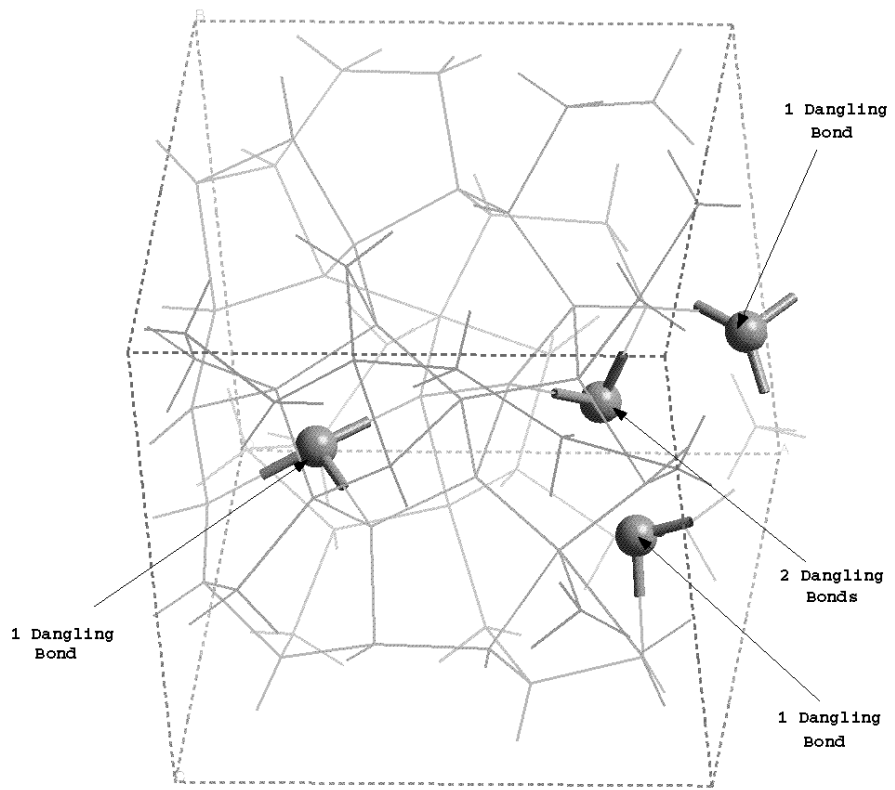
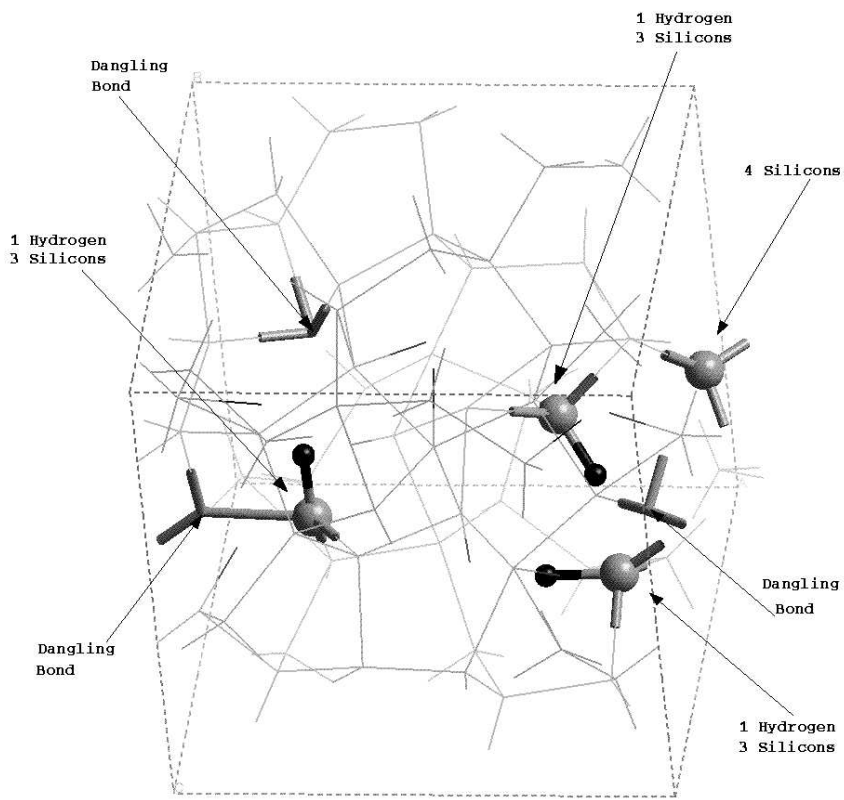
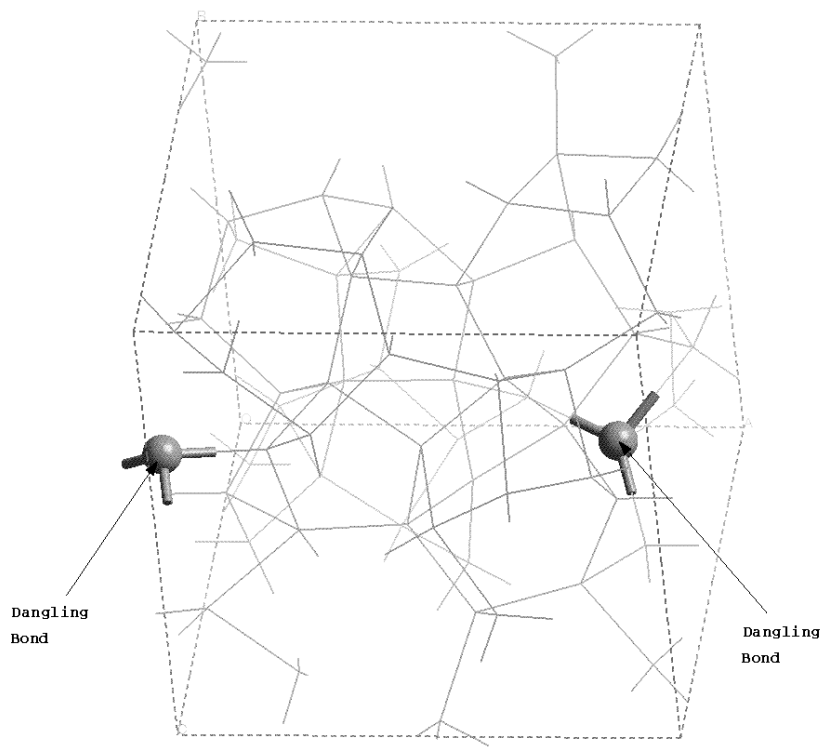
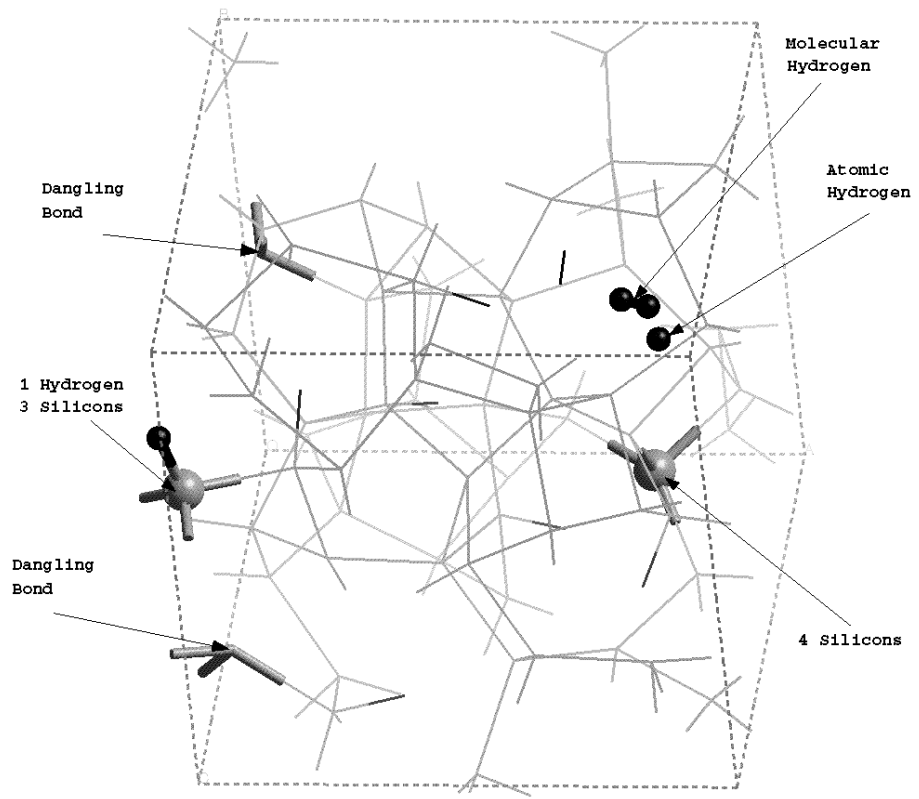
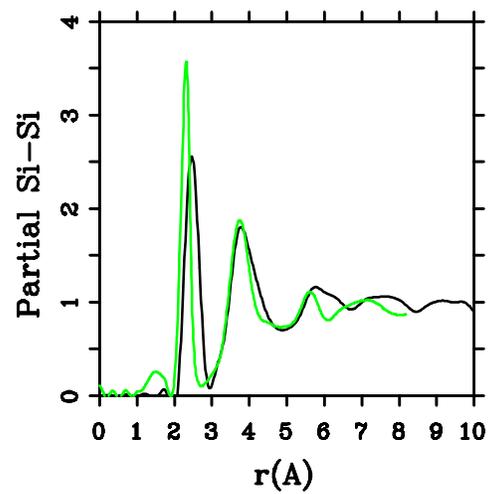
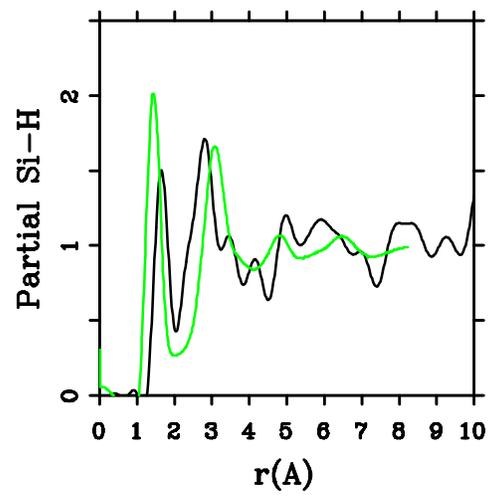
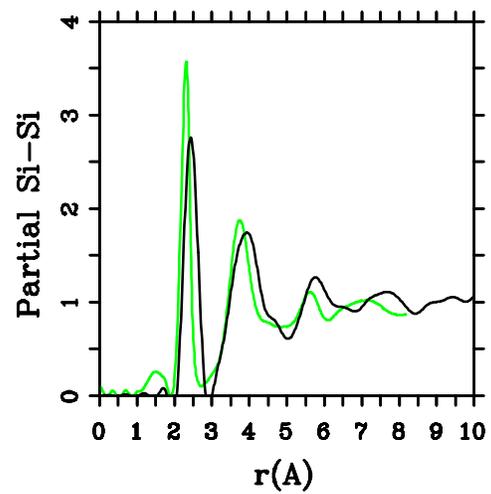
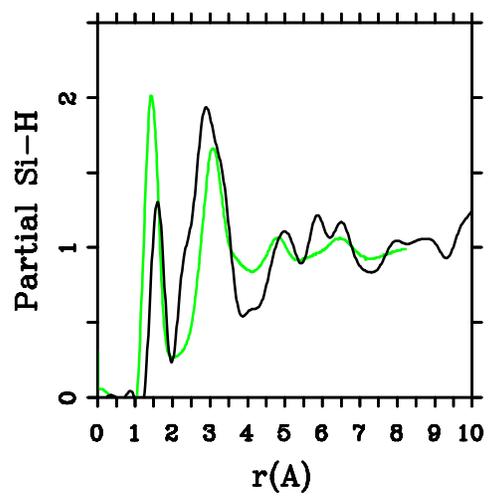
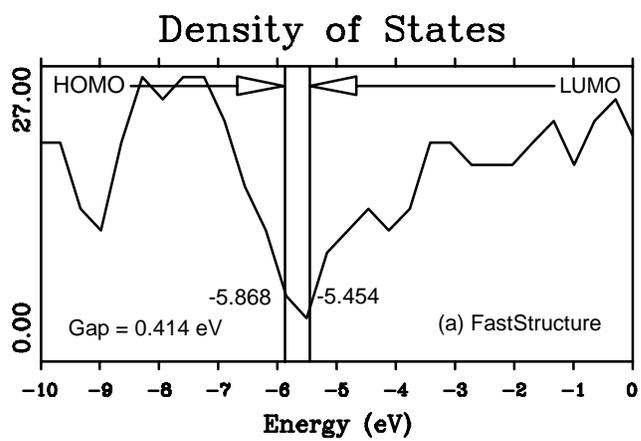
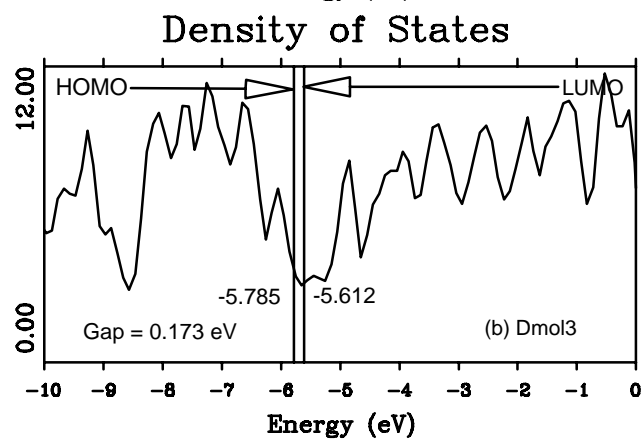
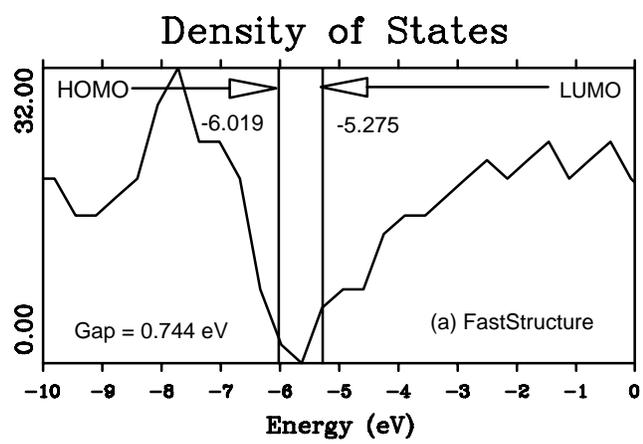
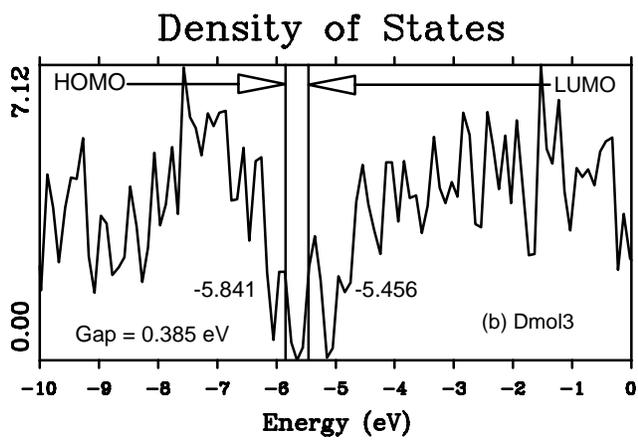
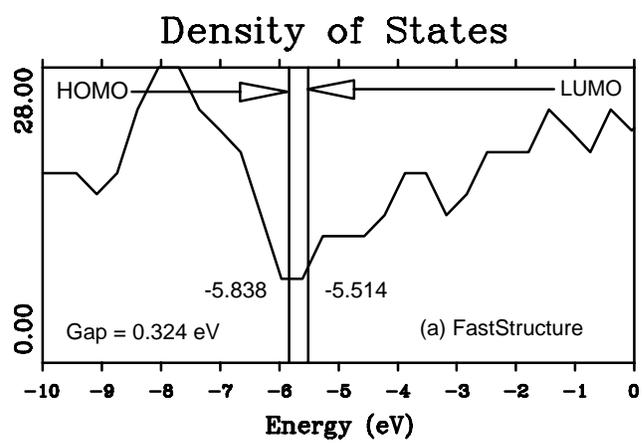
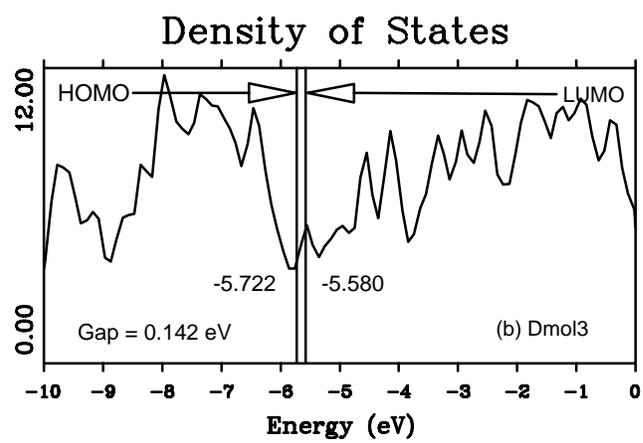
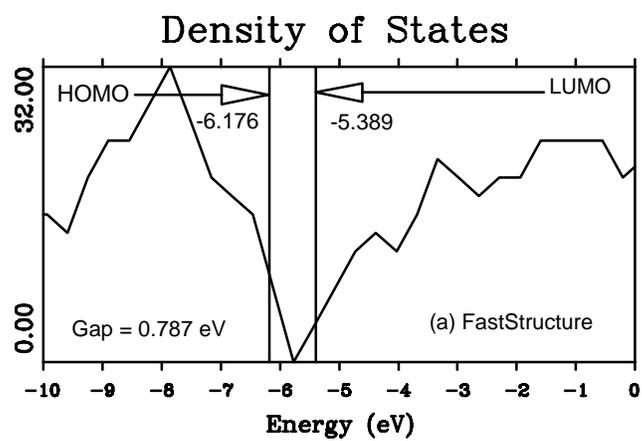
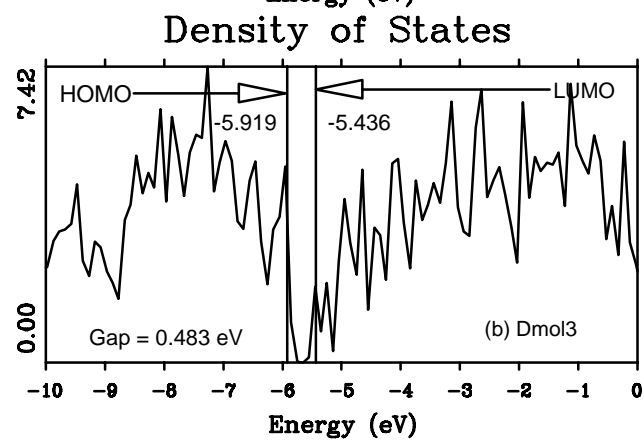
\end{document}